\documentclass[pre,twocolumn,superscriptaddress,amsmath,amssymb,endfloat]{revtex4}

\usepackage{graphicx}
\usepackage{dcolumn}
\usepackage{bm}
\usepackage[usenames,dvipsnames]{color}
\usepackage{ulem}

\begin{document}

\title{Transport and Scaling in Quenched 2D and 3D L\'evy quasicrystals}
\author{P. Buonsante}
\affiliation{Dipartimento di Fisica, Universit\`a degli Studi di Parma, Viale Usberti 7/a, I-43124 Parma, Italy}
\author{R. Burioni}
\affiliation{Dipartimento di Fisica, Universit\`a degli Studi di Parma, Viale Usberti 7/a, I-43124 Parma, Italy}
\affiliation{INFN, Gruppo Collegato di Parma, viale G. P. Usberti 7/A, 43100 Parma, Italy}
\author{A. Vezzani}
\affiliation{Centro S3, CNR--Istituto di Nanoscienze, via Campi 213A, 41125 Modena, Italy}
\affiliation{Dipartimento di Fisica, Universit\`a degli Studi di Parma, Viale Usberti 7/a, I-43124 Parma, Italy}

\begin{abstract}
 We consider  correlated L\'evy walks on a class of two- and three-dimensional deterministic self-similar structures, with correlation between steps induced by the geometrical distribution of regions, featuring different diffusion properties.  We introduce a geometric parameter $\alpha$, playing a role analogous to the exponent characterizing the step-length distribution in random systems. By a {\it single-long jump} approximation, we analytically determine the long-time asymptotic behaviour of the moments of the probability distribution,  as a function of $\alpha$  and of the dynamic exponent $z$ associated to the scaling length of the process. We show that our scaling analysis also applies to experimentally relevant quantities such as escape-time and transmission probabilities.
 Extensive numerical simulations corroborate our results which, in general, are different from those pertaining to  uncorrelated L\'evy-walks models. 

\end{abstract}

\maketitle

\section{Introduction}

L\'evy-like motions represent an important family of random motions, generated by a stochastic process with stationary and independent increments. Brownian motion constitutes a particular case of this family, but its mathematical tractability and its remarkable statistical properties have led it to become the principal model of random motion. However, different types of L\'evy  motions  have been extensively investigated, both theoretically and experimentally, as they have been found to be ubiquitous in nature: in biology  \cite{Sokolov_PRL_79_857,Lomholt_PRL_95_260603,Lomholt_PNAS_106_8204,Caspi_PRL_85_5655}, chaotic dynamics \cite{Geisel_PRL_54_616,Shlesinger_Nature_363_31}, economics \cite{Bouchaud}, search strategies \cite{Lomholt_PNAS_105_11055,Benichou_RevModPhys_83_81}, the L\'evy motion that have been observed are characterized by increments of arbitrary length $l$, with a step length distribution featuring an algebric tail $\sim l^{-(1+\alpha)}$. Such a distribution is said to be heavy-tailed and  has a diverging variance for  $\alpha< 2$. 

An interesting experimental situation where L\'evy motion can be detected is diffusion in heterogeneous and porous materials, composed of two o more types of regions with different diffusion properties. In that case, the motion of particles  consists of a sequence of scattering events occurring in the hard-scattering part of the material, followed by  long jumps  performed  at almost constant velocity in the non-scattering regions. If the material is very heterogeneous on all scales, the step length results to be  heavy-tail L\'evy-distributed. 
This description applies to transmission of light through clouds \cite{Davis_JAtmSc_59_2713},  tracer transport in heterogeneous aquifers \cite{Benson_TPM_42_211}, molecular diffusion at low pressure in porous media -- which is dominated by collision with pore walls, with ballistic motion inside the large pores \cite{Levitz_EPL_39_593} --, as well as in fractured and heterogeneous porous media \cite{Palombo_arXiv_1102_2149}. In addition, recent experiments on new disordered optical materials, the L\'evy Glasses, paved the way to the engineering of Levy-distributed step-lengths \cite{Barthelemy_Nature_453_495}. 
This phenomenology is often modeled in terms of {\it annealed} L\'evy walks, where  
step-lengths are uncorrelated \cite{Blumen_PRA_40_3964,Klafter_PhysicaA_168_637}. 

However, a key signature of these processes is that the steps are in principle not independent, as they are correlated by their mutual positions in the sample. A walker that has just traversed a large hole has a high probability of being backscattered at the following step and thus to perform a jump of roughly the same length. The step length distribution represents therefore a "quenched disorder", a standard definition in statistical mechanics of disordered systems \cite{Mezard}. Now, while the case of L\'evy walks with uncorrelated jumps is well understood \cite{Blumen_PRA_40_3964,Klafter_PhysicaA_168_637,Zumofen_PRE_47_851,Zoia_PRE_76_021116}, the correlation effects, which are expected to exert a deep influence on the diffusion properties, are still to be characterised, and the upper critical dimension above which a description in terms of uncorrelated Levy walks proves sufficient is still under debate \cite{Fogedby_PRL_73_2517,Kutner_JPA_31_2603,Schulz_PLA_298_105,Barthelemy_PRE_82_011101}. 

A first step in this direction has been taken, and to this end, quenched L\'evy processes have been studied on one-dimensional systems \cite{Barkai_PRE_61_1164,Beenakker_PRB_79_024204}. There, the effects of geometry-induced step-length correlation  on the asymptotic behavior of the mean square displacement  has been investigated in detail. More recently, again for one-dimensional systems, different aspects regarding the scaling properties of random-walk distributions, the relations between the dynamical exponents and the possible average procedures have been discussed in a common framework. This   evidenced the strong effects of correlations also on quantities averaged over all the starting sites \cite{Burioni_PRE_81_060101,Burioni_PRE_81_11127,Vezzani_PhilMag_91_1987}. Obviously, one-dimensional models represent simplified systems and may not compare quantitatively to real experiments, so far conducted on three dimensional systems. On the other hand, they have proven amenable to an exact analytic solution for the dynamics.
It is now important to keep track of what properties observed in one dimensional systems can be extended to higher dimensions. The key point is to understand the effects of step-length correlations and of averages over starting points, and what is needed is a proper model for the heterogeneous geometrical pattern where the correlated  L\'evy  motion takes place.

In this paper we model the transport process as a random walk on scale-invariant structures --- namely generalized two- and and three-dimensional Sierpinski carpets --- whose solid and empty regions represent portions of material characterized by different transmission properties. In order to reproduce a realistic propagation, our process consists of a simple diffusive random walk on the solid regions of the structure, and of a ballistic motion across its  empty regions. Therefore, while the distribution of allowed jump lengths is scale-invariant, any point in the two- or three-dimensional region of space occupied by the structure is  accessible to the walker. This is at variance with the well-known and studied case of standard diffusion on self-similar structures, where the empty regions are not accessible \cite{Ben-Avraham}, and the motion takes place on a truly fractal space.  
From a more rigorous point of view the ballistic propagation of the walker across the ``empty'' regions of fractal structure can be equivalently obtained by imposing a scale invariant pattern of persistence on the sites of a regular Euclidean lattice \cite{weiss}.

It should be remarked that the available experimental realizations are characterized by a strong amount of disorder \cite{Davis_JAtmSc_59_2713,Benson_TPM_42_211,Levitz_EPL_39_593,Palombo_arXiv_1102_2149,Barthelemy_Nature_453_495}, which might call into question our modeling  of the pattern of inhomogeneity  in terms of deterministic, self-similar structures.
However, the regular distribution of the inhomogeneities allows us to take into account several crucial points in a more controlled way. We are able to give
an estimate of the effects of the tail in  the step length distribution in terms of the geometrical parameters characterizing the L\'evy quasicrystals, and we can tune it arbitrarily to study its effects;
 we can correctly control the effects of the averages over the starting sites, which is known to have a deep influence the asymptotic properties of the mean square displacement in one-dimensional structures; most importantly, we are able to take a first step towards the comprehension of correlated L\'evy  walks in higher dimensions.

One of our main results is that, for this type of "topological'' correlations, quenched  L\'evy walks lead to a set of exponents for the asymptotic behaviors of the mean square displacements and transmission which differ from the uncorrelated case \cite{Fogedby_PRL_73_2517,Kutner_JPA_31_2603,Schulz_PLA_298_105}.  
  
In detail, in this paper, we tackle the problem by verifying the  scaling hypothesis for the probability distribution $P_j(r,t)$ , namely the probability for the L\'evy  walker to be at a distance $r$ from its  starting site $j$ at time $t$ \cite{Cates_JP_46_1059} and by evaluating the dynamical exponent $z$  associated to the growth of the scaling length $\ell(t)\sim t^{1/z}$ of the process. This is  a function of a tunable geometrical parameter, $\alpha$, describing the self similar pattern of empty regions in the L\'evy quasicrystals, and playing a role analogous to the exponent characterizing the step-length distribution in random systems. Then, by making the reasonable and well verified ``single long jump'' hypothesis \cite{Burioni_PRE_81_11127}, we estimate the tails of $P(r,t)$, i.e. the average of $P_j(r,t)$ over all the possible starting sites.  We derive an analytic  expression for the asymptotic behaviors of the moments $\langle r^p (t) \rangle$ of $P(r,t)$,  evidencing the presence of strongly anomalous diffusion \cite{Castiglione_PhysicaD_134_75}, i.e.  $\langle r^p (t) \rangle \neq C \ell(t)^p$. In the attempt of making a direct contact with experiments, we also derive the scaling properties of  exit times, as well as of the time-resolved transmission probability and transmission profiles through a slab of thickness $L$. Extensive numerical simulations are in very good agreement with the predicted behaviors.

The paper  is organized as follows: 
in the following section we describe our  Levy quasicrystals based on generalized  Sierpinski carpets (SC).
In section \ref{physical} we introduce the relevant physical quantities that we will study in this paper. In Section \ref{scalingS} we discuss the scaling hypothesis and the {\it single-long-jump approximation} which allows us to evaluate the momenta of the distribution $P(r,t)$ evidencing the presence of strongly anomalous diffusion \cite{Castiglione_PhysicaD_134_75}. In the last part of the paper  we present extensive numerical simulations proving the reliability of our scaling hypothesis and single long jump approach. In particular,  we evaluate the dynamical exponent $z$ as a function of the dimensionality, of the dynamics and of $\alpha$, a simple parameter describing the topology of the structure. For a slab of thickness $L$ we evaluate exit times, time resolved transmissions and  transmission profiles, such a results could be useful for a comparison with experiment \cite{Barthelemy_Nature_453_495}.  Section \ref{conclusion} contains our conclusions and perspectives.

\section{Structures}
\label{structures}
\begin{figure}[t!]
\includegraphics[width=7cm]{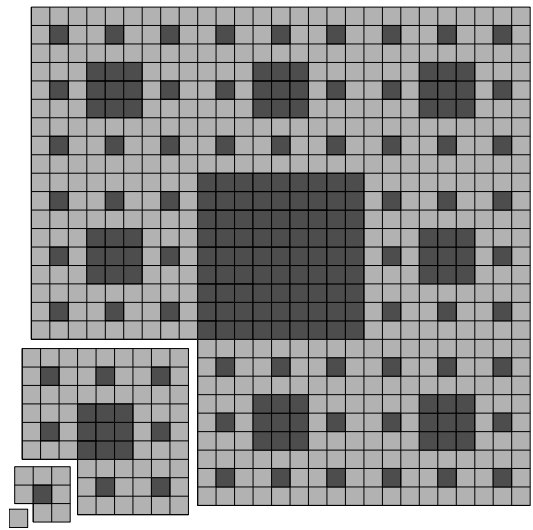}
\caption{\label{SC_3_8} Third generation of a SC with $n_u=3$ and $n_r=8$. Light squares represents actual sites of the structure. Dark squares belong to empty regions. The ``exploded view'' in the lower left corner highlights the fact that this structure is made of $n_r=8$ SCs of generation $g=2$, and so on.}
\end{figure}

In order to to realize a Levy walk on a quenched structure 
we consider a regular lattice in which empty regions have been created by removing a subset of sites. We then consider a random walker that can jump across empty regions. 
This is actually a L\'evy process provided that the linear size of such regions --- and hence the length of the ballistic jumps --- is L\'evy-distributed. 
Specifically, it is a  L\'evy walk if all jumps are performed at constant velocity, and hence  completed in a time proportional to their length.
The Cantor lattices considered in Ref.~\cite{Burioni_PRE_81_11127}, are deterministic one-dimensional structures exhibiting the desired properties. The natural two-dimensional generalization of such structures is provided by the so-called   Sierpinski carpets discussed in Ref.~\cite{Gefen_PRL_45_855}.

The self-similar character of a SC can be illustrated by observing that 
several identical objects can be assembled to give a larger SC. One way to
describe this constructive procedure is the following: arrange $n_u^2$ SC
of generation $g$ in a square array, and subsequently remove some, so 
that only $n_r<n_u^2$ of them remain. The resulting object is a SC of generation
$g+1$. If the same procedure is repeated using SCs of generation $g+1$ as
building blocks, a structure of generation $g+2$ is obtained.
Figure \ref{SC_3_8} illustrates this algorithm in the case of the standard two dimensional SC, where $n_u=3$, $n_r=8$ and the structure of generation $g=0$ is a simple lattice site. 
At each generation a $3\times 3$ array of  carpets is laid out,
and the central structure is removed, leaving a square empty region. 
Note that the linear size of the largest
empty region in the structure grows by a factor $n_u$ as the generation is increased.
A different structure, corresponding to $n_u=4$ and $n_r=12$, can be obtained by laying out a $4\times 4$ array of building blocks, and subsequently removing the four innermost structures.
In general a SC of generation $g$ features $n_r^{g-k-1}$  empty regions of linear  $n_u^k$, with $k=0,1,\cdots,g-1$. If $g$ is sufficiently large, the linear size of the empty regions spans several orders of magnitude. For instance the 3rd generation structure in Fig \ref{SC_3_8} features one empty region of linear size $9$, $8$  empty regions of linear size $3$ and $64$ empty regions of linear size $1$.

One of the parameters characterizing a SC is its fractal dimension, i.e. the average number of sites per unit surface, which is has a simple expression in terms of the constructive parameters 
\begin{equation}
d_{\rm f} = \frac{\log n_r}{\log n_u}
\end{equation}
A  more relevant characterization  for our purposes can be given in terms of the probability for the length of the jumps that can be taken at a random site of the structure. For each of the possible jump lengths $n_u^k$ this probability is given by density of sites on the boundary of empty regions of linear size $n_u^k$. It is easy to prove that the moments  of such probability 
\begin{equation}
\overline{R^p} = \lim_{g\to \infty} \sum_{k=0}^{g-1} \left(n_u^k\right)^p \frac{n_r^{g-k-1} n_u^k}{n_r^g} 
\end{equation}
converge only if $p<\alpha$, where 
\begin{equation}
\label{alpha}
\alpha = \frac{\log n_r}{\log n_u}-1.
\end{equation}
This parameter has exactly the same role as the exponent $\alpha$ characterizing the distribution of the step lengths $s$ of a Levy walk in $d$ dimensions, whose $p$-th moment
\begin{equation}
\overline{s^p} \propto \int_{s_0}^\infty ds s^{d-1} \frac{1}{s^{\alpha +d}} s^p 
\end{equation}
converges only if $p<\alpha$.

For the Sierpinski carpets discussed in Ref.~\cite{Gefen_PRL_45_855} the threshold in Eq.~\eqref{alpha}  is connected to the fractal dimension of the structure, $d_{\rm f} = \alpha+1 < 2$. Since $\alpha<1$ the size of the average empty region, i.e. the first moment of the distribution, is infinite.

\begin{figure}[b!]
\includegraphics[width=8cm]{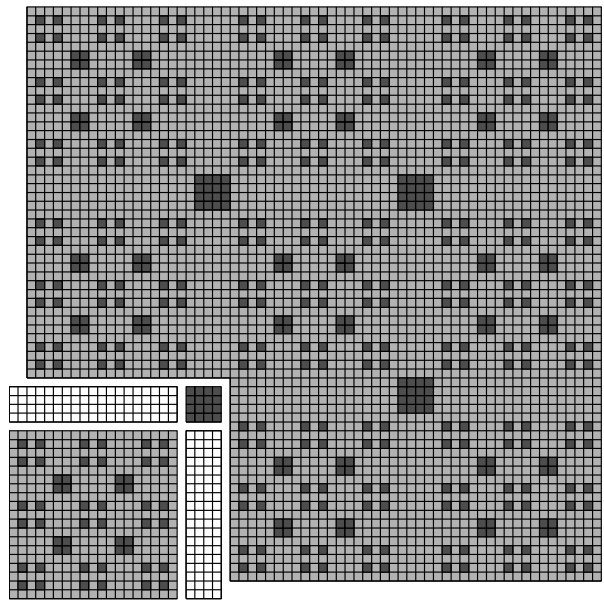}
\caption{\label{FC_9_2} Third generation of a FC with $n_r=9$ and $n_u=2$. The ``exploded view'' in the lower left corner illustrates the tiling pattern underlying the structure. Three different types of tiles are necessary to build the structure: $n_r=\nu_r^2=9$ FCs of generation $g=2$, whose side comprises $19$ sites, $(\nu_r-1)^2=4$ empty square tiles of side $\nu^2=4$ and $2(\nu_r-1)\nu_r = 12$ rectangular padding tiles, highlighted in a lighter shading.}
\end{figure}

The two dimensional counterparts of the {\it Cantor-Smith-Volterra} sets discussed e.g. in Ref.~\cite{Burioni_PRE_81_11127} are hyerarchical structures similar to Sierpinski carpets for which $\alpha>1$ and $d_f=2$. We refer to these structures as {\it fat carpets} (FC).

As we discuss above, a SC of generation $g$ can be seen as a $n_u \times n_u$ square tiling whose elements have the same size as SC of generation $g-1$. Some of these elements are actual SCs, while others are left empty.
The tiling in a FC is sligthly more complex, in that it requires three kinds of tiles, as it is illustrated in Fig.~\ref{FC_9_2}. This comes about because, in order to attain $\alpha>1$, the largest empty tiles at a given generation must grow less than the entire structure as the generation is increased. More specifically, a FC of generation $g+1$ contains $n_r = \nu_r^2$ FCs of generation $g$, whose side is $L_{\rm C}^{(g)}$, $(\nu_r-1)^2$ empty square tiles of side $L_{\rm E}^{(g)}$ and $2\nu_r(\nu_r-1)$ rectangular  $L_{\rm C}^{(g)}$ by $L_{\rm E}^{(g)}$ tiles. If $L_{\rm C}^{(1)}=0$ and $L_{\rm E}^{(g)}= n_u^g$, with $n_u< \nu_r$, we get
\begin{equation}
L_{\rm C}^{(g)} = \nu^g+(\nu-1)\frac{\nu^g-h^g}{\nu-h} \sim \nu^g \frac{2\nu-h-1}{\nu-h} 
\end{equation}
so that the fraction of sites contained in the rectangular padding tiles decreases as $n_u^g \nu_r^{-g}$. Thus, for sufficiently large generations, the contribution of the padding tiles becomes negligible and the total number of sites in the structure is proportional $n_r^g$. This means that  the threshold for finite moments in the probability for the length of the jump at a random site of the FC is once again given by the quantity in Eq.~\eqref{alpha}. Since $\nu_r = \sqrt {n_r} > n_u$, $\alpha>1$ and the average size of the empty regions is finite. All of the above discussion can be easily generalized to arbitrary dimension $d$. In fact, in  Section~\ref{scalingS}, the  {\it single-long-jump} picture is illustrated for a generic dimension $d$. As to the simulations, we mainly focus on two-dimensional structures, but we also present some results for three-dimensional {\it sponge-like} structures.

\begin{figure}
\begin{tabular}{cc}
\includegraphics[width=4cm, bb = 135 290 475 560, clip]{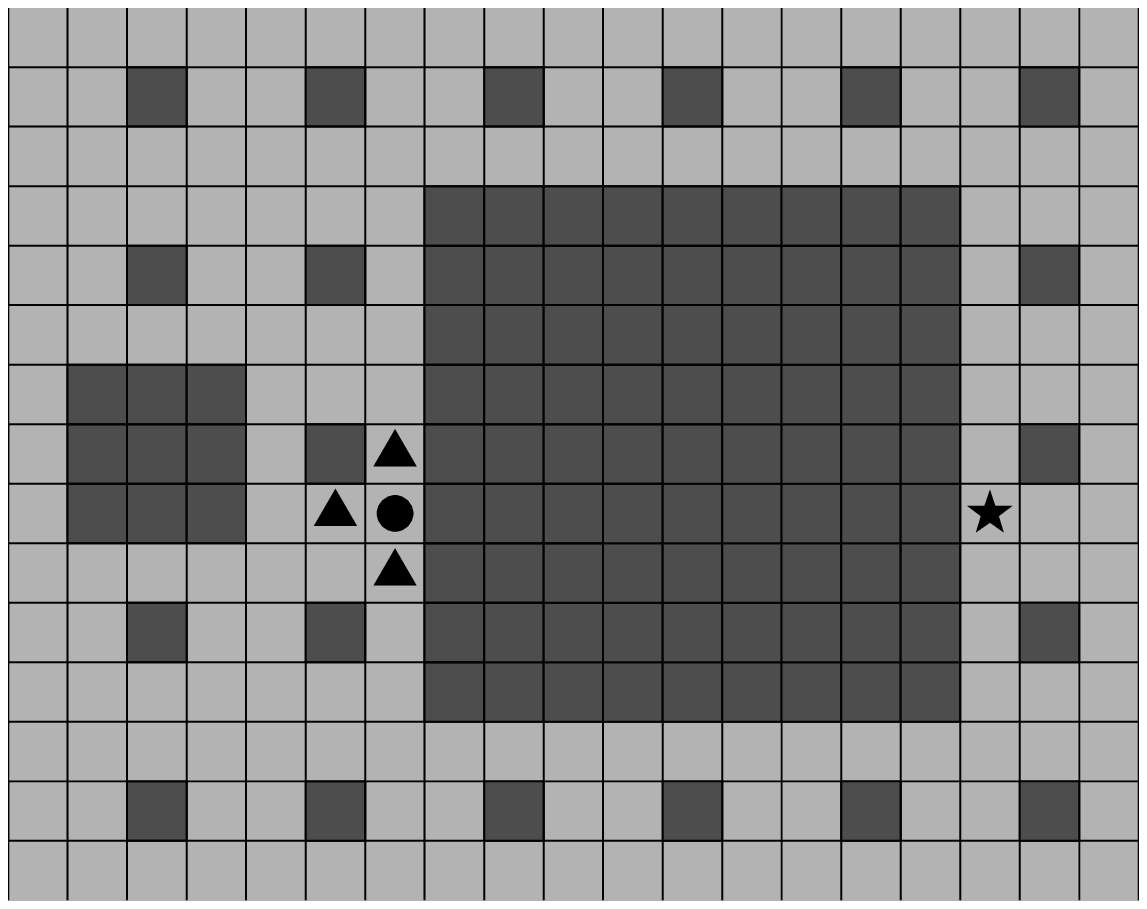} &
\includegraphics[width=4cm, bb = 135 290 475 560, clip]{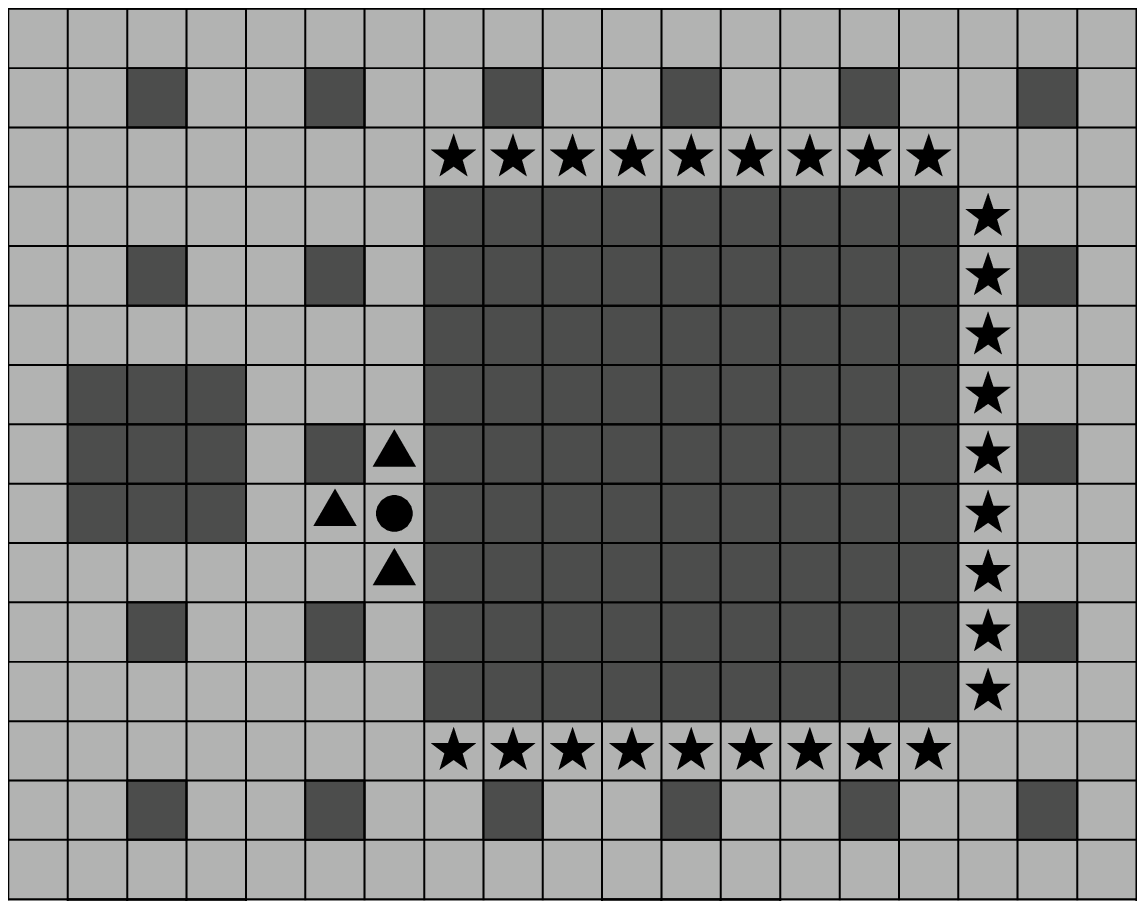}
\end{tabular}
\caption{\label{jumps} Possible jumps of a random walk initially at a site on the boundary of an empty region (black circle). The random walk can attempt a jump in one of the four possible directions, with equal probabilities. If the chosen direction points to one of the three neighbouring sites (black triangles), the jump is taken and the time is increased by one unit. If the chosen direction points to the empty region, the random traverses it. The two panels illustrate the two possible schemes for the choice of the arrival site (black pentagram). Left: the arrival site is the one facing the current position of the walker ({\it head-on dynamics}). Right: the arrival site is equiprobably chosen in the boundaries of the empty region not containing the current position of the random walk ({\it fan-out dynamics}). In any case, the time is increased by the length of the jump, as measured in units of the lattice spacing.}
\end{figure}

Let us now get back to the diffusion dynamics of the random walk. This is based on the standard dynamics on a regular square lattice: one direction out of the four possible ones pointing to the adjacent sites is randomly chosen. If a site is found in that direction at a distance of a single lattice spacing, then the jump is taken, and the time is increased by one unit. If the same jump would land inside an empty region, then the random walk traverses the entire region and lands at a site belonging to a different boundary from that containing the starting site. Fig.~\ref{jumps} illustrates the  two possible schemes for the choice of the arrival site we consider in this work. In the simpler scheme (left panel) the walker lands at the site facing its current position ({\it head-on dynamics}). In the second scheme (right panel) the landing site is randomly chosen among those belonging to any of the sides of the empty region except the one currently hosting the random walk ({\it fan-out dynamics}). In any case, since the random walk moves at constant velocity, the time required for a jump equals its length.

Note that the jump probability of these two dynamics are different at any distance scale. Therefore we expect that parameters usually exhibiting universal character might be different for the two cases.
In fact, our simulations evidence that  the dynamical exponent governing the growth of the characteristic length of the process  seems to depend on the dynamics, on the dimensionality of the system and on the geometric parameter $\alpha$. Conversely, a very robust feature we observe is that anomalous diffusion takes place strictly for $\alpha<1$ for every case considered here. This is at variance with the case of uncorrelated L\'evy walks, where $\alpha<1$ and $1<\alpha<2$ correspond to purely ballistic propagation and superdiffusion, respectively.

\section{Physical quantities}
\label{physical}
The diffusion process on the structures under concern can be characterized by several quantities. One of the most natural is the probability $P_j(r,t)$ that  at time $t$ the random walk is at a distance $r$ from the lattice site $j$ it started from. The average of this probability over all the lattice sites,
\begin{equation}
\label{avgP}
P(r,t) = \frac{1}{N}\sum_j P_j(r,t),
\end{equation}
where $N$ is the number of lattice sites, is a meaningful quantity as well, in view of the spatial inhomogeneity of the SCs and FCs \cite{average}. As we will discuss shortly, it highlights the anomalies in the behaviour of the diffusive process.
These probability distributions are completely defined by their moments,
\begin{equation}
\label{mom}
\langle r^p_j(t)\rangle \!=\! \int dr\,r^p\,  P_j(r,t), \quad \langle r^p(t)\rangle \!=\! \int dr\,r^p\,  P(r,t)
\end{equation} 

A further  quantity worth analyzing is the escape time probability, namely the probability ${\cal E}(L,t)$ that a walker starting at a random site of a structure of linear size $L$ escapes from it at time $t$.

An even more direct contact with the experiment is provided by the {\it conditional first-passage time} $T_{j k}(L,t)$, i.e. the probability that a particle injected at a site $j$  belonging to one boundary of the structure  hits a site $k$ belonging to the opposite boundary at time $t$ before going back to the boundary containing $j$, which is therefore absorbing. 
 This means that there is a (large) probability that the  injected particles are backscattered by the structure. As to the direction parallel to the entrance and exit boundaries, we consider two possible schemes. In the first, the random walk is simply reflected by the boundaries of the structure. In the second the environment seen by the random walk is periodic, i.e. it consists of an infinite stack of identical structures of some fixed generation. 

The conditional first-passage time gives access to experimentally relevant quantities. For instance, it is clearly related, through integration over time, to  the transmission profiles shown in Ref.~\cite{Barthelemy_Nature_453_495},
i.e. the average number of processes transmitted through the sample at a transverse distance $r$ from the injection point. Specifically, on a  2D system of linear size $L$, these are obtained as 
\begin{equation}
\label{trP}
{\cal T}_{\rm p}(r,L) = \frac{1}{L} \sum_{j} \int_0^\infty dt T_{j\, j+r}(L,t)
\end{equation}

 Likewise, averaging over entrance sites and summing over exit sites gives the the time-resolved transmission probability,
\begin{equation}
\label{trTi}
{\cal T}(L,t) = \frac{1}{L} \sum_{j k} T_{j k}(L,t).
\end{equation}
Finally, integrating this quantity over time gives the total transmitted intensity, i.e. the fraction of processes transmitted through the sample 
\begin{equation}
\label{Ti}
{\cal I}(L) =\int_0^\infty dt {\cal T}(L,t).
\end{equation}

\section{Scaling analysis and  {\it Single long jump} ansatz}
\label{scalingS}
Several of the quantities discussed in the previous Section depend on time and on a spatial variable. The latter can be the walk distance $r$ in the case of the local and average probabilities $P_j(r,t)$ and $P(r,t)$, or the system size $L$ in the case of the escape probability.
A generic function of space and time is expected to exhibit the following scaling structure
\begin{equation}
\label{scaling}
f(r,t) = t^\gamma \tilde f\left(\frac{r}{\ell(t)}\right),
\end{equation}
where the asymptotic behaviour of the characteristic length is
\begin{equation}
\label{ell}
\ell(t) \sim t^\frac{1}{z}
\end{equation}
Notice that the prefactor in r.h.s. of Eq.~\eqref{scaling} depends only on time. An alternative, equivalent form of this equation is also possible, featuring a prefactor depending on the spatial variable alone, $r^{\gamma z}$, and a scaling function simply related to $\tilde f$.

The exponent $\gamma$  in the prefactor can be determined from  known features of the quantity under concern. For instance, the expected dependence of the return probability on the spectral dimension $d_{\rm s}$ of the structure, $P(0,t)\sim t^{-d_{\rm s}/2}$ \cite{Alexander_JPL_43_L625,Hattori_PTP_92_108}, entails that $\gamma = d_{\rm s}/2$ for the local probability $P_j(r,t)$. Furthermore,  the time-independence of the norm of $P_j(r,t)$ yields $z = 2 d/d_{\rm s}$, where $d$ is the dimensionality of the system.

As to the function $\tilde f(x)$, one expects it to decay very rapidly unless, at time $t$, the random walk was able to take jumps much longer than $\ell(t)$.
The rapid decay is typical of local quantities, whereas average quantities usually exhibit an anomalous behaviour.
We illustrate the origin of the slow decay  by focusing on  $P_j(r,t)$ and its average $P(r,t)$, defined in Eq.~\eqref{avgP}. A slow decay of $\tilde f(x)$ is expected when the random walk is able to take arbitrarily large jumps at arbitrarily small times. This is precisely the case of $P(r,t)$, since the sum in  Eq.~\eqref{avgP} includes starting sites in the vicinity of empty regions of arbitrary size. 

The situation is somewhat different for the local probability $P_j(r,t)$. It is indeed true that the starting site can be close to an empty region of very large linear size, say $\bar L$, so that a jump of length $\bar L$ can occur at relatively early times. However, this merely affects the transient part of the dynamics. After a suitably long time $\ell(t)\gg \bar L$, the regular behaviour is recovered, since empty regions much wider than $\ell(t)$ are far removed from the position of the random walk. For a similar reason we expect the long-time asymptotic behavior of $P_j(r,t)$ to be independent of $j$. Indeed for $\ell(t)$ much larger than the distance between sites $i$ and $j$ we expect that $P_j(r,t) \simeq P_i(r,t)$.

The fast or slow decay of $\tilde f(x)$ is mirrored by the asymptotic behaviour of the moments defined in Eq.~\eqref{mom}. The rapidly decaying scaling function in the local probability indeed yields
\begin{equation}
\label{regularm}
\langle r_j(t)\rangle \sim \sqrt[p]{\langle r_j^p(t)\rangle} \sim \ell(t)
\end{equation}
independent of the starting point. 

The expected more complex behaviour of the average moments $\langle r^p(t)\rangle$ when slow decay is present can be studied in the {\it single-long-jump } picture \cite{Burioni_PRE_81_060101,Burioni_PRE_81_11127,Vezzani_PhilMag_91_1987}. This corresponds to  assuming that the average probability can be split into two contributions, a regular, rapidly decaying part $P^{({\rm r})}(r,t)$ and an anomalous tail  $P^{({\rm a})}(r,t)$. The  latter is assumed to be generated by a single jump, much larger than the characteristic length, and it is expected to give rise to a significant contribution in the region $r \gg \ell(t)$ where the former is negligible.

In order to estimate the anomalous contribution, we have to determine the probability that the random walk takes a jump of size  $\bar \ell\gg \ell(t)$ at time $t$. This is proportional to the density of regions of linear size $\bar \ell$ times the number of sites from which a random walk can reach one of such regions in a time $t$.  Recalling that the average distance covered by the walker in a time $t$ is given by $\ell(t)$, the sites accessible to a single region of linear size $\bar \ell$ are enclosed in a region whose volume is proportional to $ \ell(t)\,\bar \ell^{d-1}$. The density of sites corresponding to  a volume of linear size $\ell(t)$ is $\ell(t)^{d_{\rm f}-d}$, where $d_{\rm f}$ is the fractal dimension discussed in Section~\ref{structures}. Observing that the empty regions in the self similar structures thereby discussed have discrete linear sizes $\bar \ell = n_u^k$ and discrete densities $n_r^{-k}$, we estimate the anomalous contribution to the average probability as
\begin{equation}
\label{Pa}
P^{({\rm a})}(r,t)= \ell(t)^\sigma \sum_{n_u^k>\ell(t)}  \left(\frac{n_u^{d-1}}{n_r}\right)^k \delta[r-{\rm min}(n_u^k,t)]
\end{equation}
where the delta function takes care of empty regions whose linear size is larger than the longest possible distance (ballistically) covered by the random walker in a time $t$, and
\begin{equation}
\sigma = d_{\rm f} - d+1 = \left\{
\begin{array}{cc}
\alpha & {\rm if}\,\alpha<1 \\
1 & {\rm if}\,\alpha \geq 1
\end{array}
\right. .
\end{equation}
We can now analyze the generic moment of the average probability.
According to its definition, the regular part of such probability results in a regular contribution, like in Eq.~\eqref{regularm}: $\langle r^p\rangle^{({\rm r})}\sim \ell(t)^p$. The contribution from $P^{({\rm a})}$ can be evaluated as
\begin{eqnarray}
\langle r^p \rangle^{({\rm a})} &=& \ell(t)^\sigma \sum_{\ell(t)<n_u^k<t} \left(n_u^k\right)^p   \left(\frac{n_u^{d-1}}{n_r}\right)^k \nonumber \\
&+& \ell(t)^\sigma t^p \sum_{n_u^k>t}  \left(\frac{n_u^{d-1}}{n_r}\right)^k
\end{eqnarray}
Performing the summations and joining the regular and anomalous contributions. we obtain 
\begin{equation}
\label{rp}
\langle r^p \rangle = C_1 \,t^{\frac{p}{z}} + C_2\, t^{\frac{p-\alpha+\sigma}{z}}+ C_3\, t^{p-\alpha+\frac{\sigma}{z}}
\end{equation}
where the $C_j$'s are time-independent coefficients and we made use of Eq.~\eqref{ell}. Equation~\eqref{rp} shows that the asymptotic behaviour of the generic moment of the average probability is actually more complex than that given in Eq.~\ref{regularm}, i.e. the system exhibits strongly anomalous diffusion \cite{Castiglione_PhysicaD_134_75}. More specifically on thin carpets ($\alpha < 1$) 
\begin{equation}
\langle r^p \rangle \sim
\left\{
\begin{array}{ll}
t^\frac{p}{z} & {\rm if}\,p<\alpha \\
t^{p-\alpha\frac{z-1}{z}} & {\rm if}\,p \geq \alpha
\end{array}
\right.,
\label{rpmin1}
\end{equation}
whereas on fat carpets ($\alpha \geq 1$)
\begin{equation}
\langle r^p \rangle \sim
\left\{
\begin{array}{ll}
t^\frac{p}{2} & {\rm if}\,p<2 \alpha-1 \\
t^{p-\alpha+\frac{1}{2}} & {\rm if}\,p \geq 2\alpha-1
\end{array}
\right.,
\label{rpmag1}
\end{equation}

Let us now turn to the scaling properties of the escape time probability. 
The scaling hypothesis states that ${\cal E}(L,t)$ can depend on the system size
$L$ only through the ratio $L/\ell(t)$ i.e.
\begin{equation}
\label{scalE}
{\cal E}(L,t) = L^\eta  \tilde f(L/\ell(t)).
\end{equation}
Imposing the normalization 
\begin{equation}
1= \int {\cal E}(L,t) dt = \int L^\eta  \tilde f(L/\ell(t)) dt,
\end{equation}
we get $\eta=-z$, where we once again made use of Eq.~\eqref{ell}.

Similar considerations apply to the transmission-related quantities introduced in Section \ref{physical}. The time-resolved transmission probability, Eq. \eqref{trTi}, is expected to obey a scaling law of the same form  as Eq.~\eqref{scalE},
\begin{equation}
\label{scalT}
T(t,L) = L^\eta  \tilde f(L/\ell(t)),
\end{equation}
although with different scaling function and exponent. The latter is fixed by 
recalling that the Einstein relation connecting the  growth of the characteristic  length  and the total transmitted intensity as a function of the linear size of the system \cite{Burioni_PRE_81_060101,Cates_JP_46_1059} gives  ${\cal I}(L)\sim L^{1-z}$. On the other hand, plugging  Eqs.~\eqref{ell} and \eqref{scalT} into Eq.~\eqref{Ti} gives
\begin{equation}
{\cal I}(L)=\int T(t,L) dt = L^\eta \int \tilde f(L/\ell(t)) dt \sim L^{\eta+z},
\end{equation}
which yields $\eta=1-2z$.

The transmission profile ${\cal T}_{\rm p}(r,L)$ defined in Eq.~\eqref{trP}, does not depend on time, which was integrated over, but on two distances, $r$ and $L$. The relevant scaling function is expected to depend only on the ratio of these distances,
\begin{equation}
\label{scalTp}
{\cal T}_p(r,L)=L^{\eta} \tilde f(r/L)
\end{equation}
Of course $\tilde f$ and $\eta$ are different from those appearing in Eqs.~\eqref{scalE} and \eqref{scalT}. Since the integration of ${\cal T}_{\rm p}(r,L)$ over all possible distances $r$ gives once again the total transmitted intensity defined in Eq.~\eqref{Ti}, the above-mentioned Einstein relation fixes the exponent to $\eta = -z$.

\section{Simulations and Results}
\label{simulation}
In the present section we provide numerical evidence corroborating the scaling hypothesis and the {\it single-long-jump} picture illustrated in the previous section. An effective encoding of the topology of the structure in a matrix of size $g \times L$, where $g$ and $L$ are the generation of the carpet and the length of its side, respectively, allowed us to address $L$'s up to a few million sites. In fact, after a straightforward generalization, the matrix relevant to a given carpet can be used to describe its $d$-dimensional counterpart with no additional memory demand. We will present some results for three dimensional structures as well. 

We start by focusing on the local probability density $P_j(r,t)$. After checking that, as expected, the asymptotic behaviour of this quantity does not depend on the starting site $j$, we collected extensive data on $P_O(r,t)$, where $O$ denotes one of the vertices of the structure under examination.
\begin{figure}[t!]
\includegraphics[width=\columnwidth, bb = 105 264 475 578,clip]{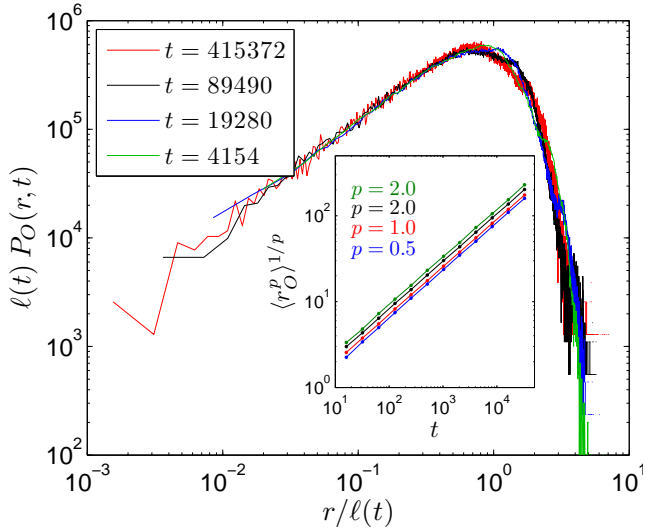}
\caption{\label{PL_carpet} (Color online) Scaling of the local probability for the (thin) SC with $n_r=8$ and $n_u=3$, corresponding to $\alpha \approx 0.8928$ ({\it head-on} dynamics). The inset provides a check of Eq.~\eqref{regularm} for some values of $p$. A fit of the data according to Eq.~\eqref{ell} gives $z^{-1}\approx 0.553$. }
\end{figure}
Figure~\ref{PL_carpet} demonstrates that $P_O(r,t)$  satisfies the scaling hypothesis in Eq.~\eqref{scaling}, with a suitable choice of the exponent governing the asymptotic growth of the characteristic length $\ell(t)$, as dictated by Eq.~\eqref{ell}. The expected steep decay of the scaling function $\tilde f(x)$ is also apparent from Fig.~\ref{PL_carpet}, while the inset in the same figure demonstrates the correctness of our surmises in Eqs.~\eqref{regularm} and \eqref{ell}. For 1D systems, analytic arguments show that the  exponent appearing in Eq.~\eqref{ell} depends on the  parameters $n_u$ and $n_r$ determining the topology of the structure through the parameter  $\alpha$ defined in Eq.~\eqref{alpha}, as $z=1+\alpha$  \cite{Burioni_PRE_81_060101}. 
\begin{figure}[t!]
\includegraphics[width=\columnwidth,bb = 105 264 475 578,clip]{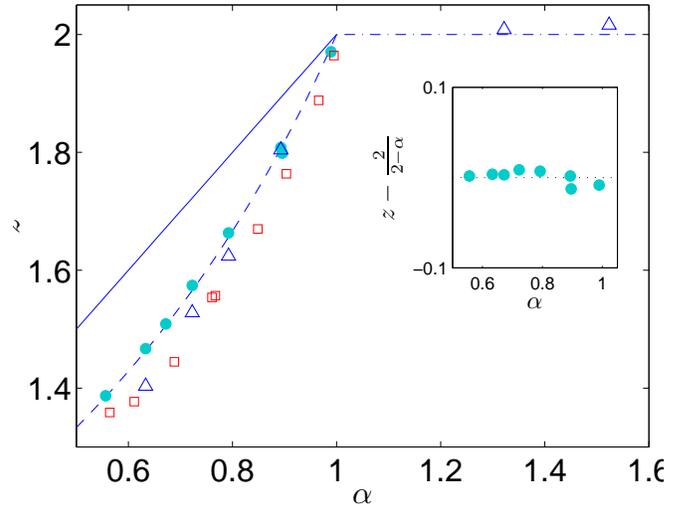}
\caption{\label{z_vs_alpha} (Color online) Dynamic exponent governing the asymptotic behaviour of the characteristic length in Eq.~\eqref{ell} for several quasicrystals and jump schemes. Head-on dynamics on 2D structures (circles); fan-out dynamics on 2D structures (triangles); head-on dynamics on 3D structures (squares). The solid line is 
 the analytically known result for the 1D case, $z=\alpha+1$,  \cite{Burioni_PRE_81_060101}, shown for comparison. The dashed line is an ansatz for the head-on dynamics on 2D systems, $z=z/(2-\alpha)$. The dash-dot line is the  result for ``ballistic'' motion.
 Note that we examined structures characterized by very different geometries yet corresponding to very similar values of the parameter $\alpha$ in Eq.~\eqref{alpha}. Specifically we considered $\alpha = \log 8/\log 3-1 \approx \log 40/\log 7 -1$ in the two-dimensional case and $\alpha = \log 3904/\log 20-2 \approx \log 218/\log 7 -2$ in the three dimensional case. }
\end{figure}
Fig.~\ref{z_vs_alpha} shows the dependence of $z$ on $\alpha$ as obtained from fits of Eq.~\eqref{regularm} for two and three dimensional structures. For the former we employed both dynamics, whereas for the latter only the {\it head-on} dynamics was considered.
 Notice that  we examined structures characterized by different values of $n_u$ and $n_r$, which combine to give very similar values of $\alpha$, and that the corresponding fitted values of $z$ exhibit a very satisfactory agreement. This is rather convincing evidence that, as in the 1D case, the exponent $z$   depends on $\alpha$ rather than on $n_u$ and $n_r$ separately.
Also, our data shows that the anomalous growth of $\ell(t)$ -- i.e. $z<2$ -- is present only for $\alpha<1$, whereas for $\alpha\geq 1$ the growth of the characteric length is consistent with $z=2$. 
That is, $\alpha <1$ and $\alpha \geq 1$  correspond to superdiffusive and  standard diffusive behaviour, respectively. This applies independent of the system dimensionality and chosen dynamics.
Finally, we notice that, for the {\it head-on} dynamics on 2D structures the dynamic exponent seems to be a very simple function of the L\'evy parameter, $z = 2/(2-\alpha)$, which might indicate that an analytic formulation is possible as in the 1D case. 

\begin{figure}[t!]
\includegraphics[width=\columnwidth, bb = 105 264 475 578,clip]{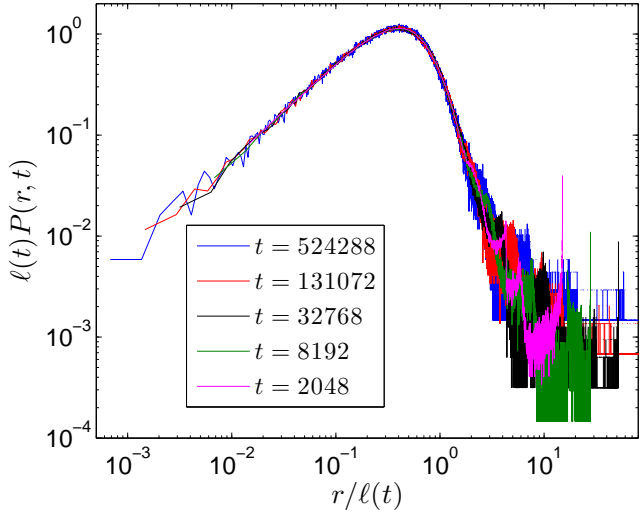}
\caption{\label{P_A_carpet} (Color online) Scaling of the average probability for the (thin) SC with $n_r=8$ and $n_u=3$, corresponding to $\alpha \approx 0.8928$ ({\it head-on dynamics}). The characteristic length $\ell(t)$ is exactly the same as that used in Fig.~\ref{PL_carpet}, $\ell(t)\sim t^{1-\alpha/2}$. }
\end{figure}

\begin{figure}[b!]
\includegraphics[width=\columnwidth, bb=100 315 475 505,clip]{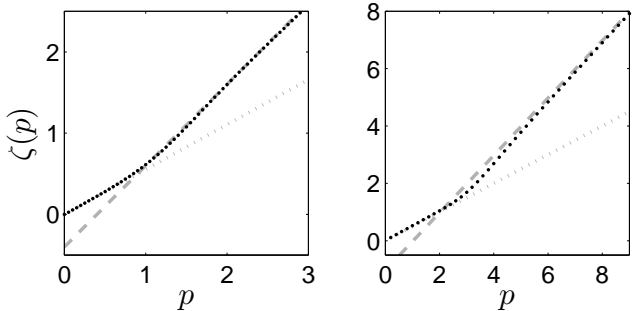}
\caption{\label{avg_mom} (Color online) Exponent governing the asymptotic behaviour  $\langle r^p\rangle = t^{\zeta(p)}$ of the moments of $P(r,t)$, Eq.~\eqref{mom}. Left: SC with $\alpha = \log 8/\log 3-1$; dotted line: $\zeta = z^{-1} p$; dashed line: $\zeta = p-z^{-1}(z-1) \alpha$; symbols: data fits. Right: FC with $\alpha = \log 16/\log 3 -1$; dotted line: $\zeta = p/2$; dashed line: $\zeta = p- \alpha+1/2$; symbols: data fits. }
\end{figure}

We now turn to the properties of the average probability in Eq.~\eqref{avgP}. Figure \ref{P_A_carpet} shows that the scaling relation in Eq.~\eqref{scaling} applies for sufficiently small values of $r/\ell(t)$, where $\ell(t)$ is the characteristic length dictating the asymptotic behaviour of the local probabilities $P_j(r,t)$. For large values of $r/\ell(t)$ it is possible to recognize structures ultimately arising from the complex superposition of delta functions appearing in Eq.~\eqref{Pa}. The long tails characterizing the distributions $P(r,t)$ give rise to the anomalous behaviour in the average moments described by Eqs~\eqref{rpmin1}-\eqref{rpmag1}. These results of the {\it single-long-jump} hypothesis are indeed confirmed by Fig.~\ref{avg_mom}, showing the fitted values of the exponent $\zeta$ dictating the asymptotic dependence of the moments on time,  $\langle r^p\rangle \sim t^{\zeta(p)}$. The linear dependence of  $\zeta(p)$ on $p$ and the change in slope predicted by Eqs~\eqref{rpmin1}-\eqref{rpmag1} is clearly recognizable. 

\begin{figure}[h!]
\includegraphics[width=\columnwidth,bb = 87 290 475 520,clip]{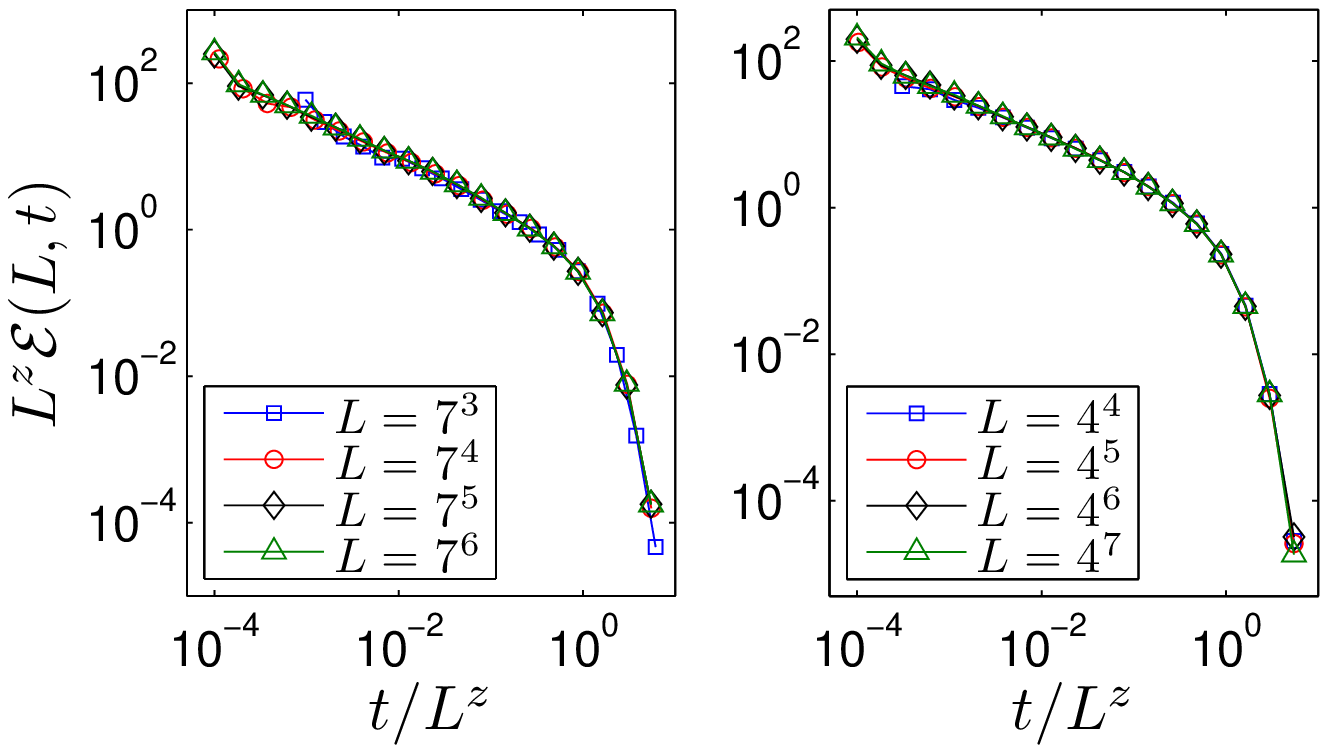}
\caption{\label{escT} (Color online) Scaling of escape time probability. Left: $\alpha = \log 24 / \log 7-1$.  Right: $\alpha = \log 12 / \log 4-1$. The numeric data have been obtained by adopting the {\it head-on} dynamics described in Fig. \ref{jumps}. }
\end{figure}

\begin{figure}[h!]
\includegraphics[width=\columnwidth,bb = 87 290 475 520,clip]{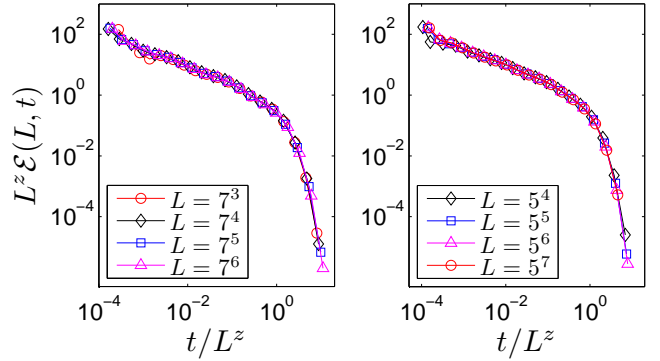}
\caption{\label{escT_DJ} (Color online) Scaling of escape time probability. Left: $\alpha = \log 24 / \log 7-1$.  Right: $\alpha = \log 16 / \log 5-1$. The numeric data have been obtained by adopting the {\it fan-out} dynamics described in Fig. \ref{jumps}. }
\end{figure}

We conclude the present section by discussing further results confirming the general validity of scaling approach.
Figs.~\ref{escT} and \ref{escT_DJ} refer to the escape-time probability for the {\it head-on} and {\it fan-out} dynamics, respectively. It is apparent that  ${\cal E}(L,t)$ obeys  Eq.~\eqref{scalE}, where $\eta=-z$ and $z$ depends on $\alpha$ according to the behavior described in Fig. \ref{z_vs_alpha}.

\begin{figure}[t!]
\includegraphics[width=\columnwidth,bb = 87 290 475 520,clip]{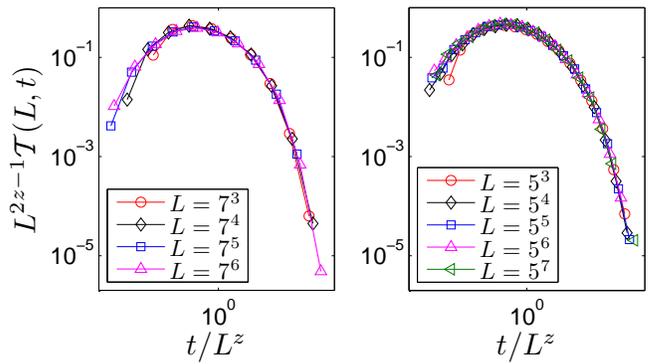}
\caption{\label{tr_tP} (Color online) Scaling of  time-resolved transmission probability. Left: $\alpha = \log 24 / \log 7-1$.  Right: $\alpha = \log 16 / \log 5-1$. The numeric data have been obtained by adopting the {\it fan-out} dynamics described in Fig. \ref{jumps}. }
\end{figure}

\begin{figure}[t!]
\includegraphics[width=\columnwidth,bb = 115 275 475 578,clip]{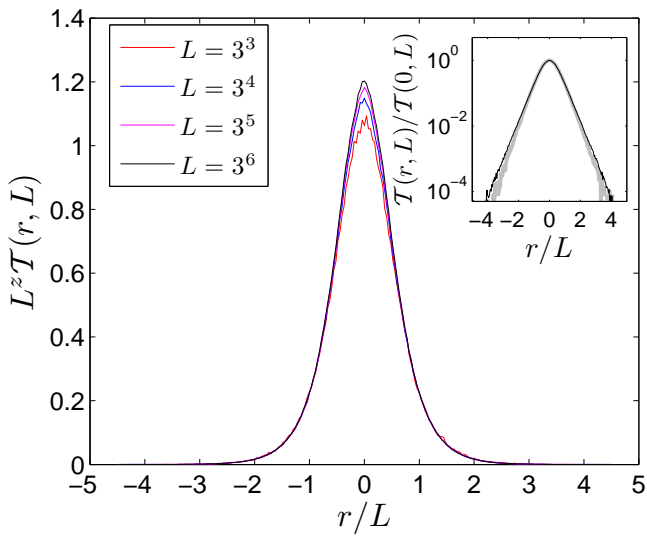}
\caption{\label{tProf} (Color online) Scaling of  the transmission profiles for the standard SC, $\alpha = \log 8 / \log 3-1$, according to Eq.~\eqref{scalTp}. The inset shows a comparison between the profile for $L=3^6$ (black thin line) and that for a homogeneous sample (gray thick line). The numeric data have been obtained by adopting the second dynamics described in Fig. \ref{jumps}. }
\end{figure}

The same applies to transmission-related quantities, as it is clear from  Figs.~\ref{tr_tP} and \ref{tProf}, which prove the validity of scaling Equations \eqref{scalT} and \eqref{scalTp}, respectively.
We note that Fig.~\ref{tProf} shows that at small $r/L$ values the convergence of the   transmission profiles is attained only for sufficiently thick samples. At first sight this could be related to the cusp-like shape observed in the experiments \cite{Barthelemy_Nature_453_495}.
However, when plotted in logarithmic scale, the profiles turn out to be exponential for large $r/L$ and bell-shaped at the center. This is the same shape characterizing the transmission profile for a homogeneous sample (see inset of Fig.~\ref{tProf}). The discrepancy between the experimental profiles and our results is very likely to be ascribed to the fact that the experimental samples are much more inhomogeneous than our L\'evy quasicrystals.

\section{Conclusions}
\label{conclusion}
In this paper we analyze L\'evy walks whose step-length distribution is characterized by correlations arising from  the structural properties of the  L\'evy quasicrystal through which the process takes place. 
One of the main outcomes of our work is that this correlation has a non-trivial  influence over all of the properties of L\'evy walks we have addressed. 
The difference between correlated and uncorrelated L\'evy walks are not a peculiarity of the one-dimensional case examined in Refs. \cite{Burioni_PRE_81_060101,Burioni_PRE_81_11127,Vezzani_PhilMag_91_1987}, but persists in two- and three dimensions, at least as long as the correlations arise from the self-similar  inhomogeneity pattern of the L\'evy quasicrystals hereby considered. 
The scaling analysis we develop shows that the dynamic exponent $z$ governing the asymptotic behaviour of the characteristic length of the local probability $P_j(r,t)$ has a wider scope than the mean-square displacement usually employed in the description the process of (super-)diffusion through the sample. 
Our {\it single-long-jump} ansatz provides a very satisfactory quantitative description of the strongly anomalous asymptotic behaviour of the process emerging after averaging over all possible starting points. 
In order to make a more direct contact with experiments we extend our scaling analyis to escape-time probabilities, time-resolved transmission probabilities and transmission profiles.
Our scaling picture  is indeed confirmed by extensive numeric simulations covering a significant range of sample sizes and heavy-tailed jump distributions.

A further important message from our results is that L\'evy quasicrystals are an ideal testbed for reproducing the superdiffusive features induced by structural properties of the sample.
In this respect the question naturally arises as to what extent  the above scenario is robust to the introduction of disorder typical of the  highly inhomogeneous structures usually invoked for the realization of correlated L\'evy walks 
\cite{Davis_JAtmSc_59_2713,Benson_TPM_42_211,Levitz_EPL_39_593, Palombo_arXiv_1102_2149, Barthelemy_Nature_453_495}. Specifically, the comparison of our transmission profiles and those reported in Ref.~\cite{Barthelemy_Nature_453_495} suggests that  L\'evy quasicrystals are much more homogeneous than the  L\'evy glass realized at {\it European Laboratory for Non-Linear Spectroscopy}. 
 We are currently working at exending our analysis to structures characterized by disorder and/or a higher degree of inhomogeneity than L\'evy quasicrystals \cite{inpreparation}.

While we deem these issues of utter interest, we think that L\'evy quasicrystals should not be regarded as a simple toy model, but as an alternate, highly controllable route to observe correlated L\'evy walks in engineered materials. 
In this perspective, the ``cleanliness'' of the samples suggested by the bell-like shape of the  transmission profile in Fig.~\ref{tProf} could even prove to be a desirable feature. Indeed, in the absence of the strong inhomogeneity characterizing   most of the relevant physical systems, the superdiffusive character of L\'evy walks could emerge more cleanly from experimental measurements. 
For instance a data-collapse of measured time-resolved transmission probabilities like in Fig.~\ref{tr_tP} would provide a measurement of the dynamical exponent characterizing the asymptotic behaviour of the process.   
 
The comparison of measurements on   L\'evy glass and  L\'evy quasicrystals would then highlight the role of strong inhomogeneity in quenched L\'evy walks.

\begin{acknowledgments}
This work has been partially supported by the MIUR Project 
{\it P.R.I.N. 2008} ``Nonlinearity and disorder in classical and
quantum processes.''
Some of our simulations have been performed on the pc cluster {\it Turing} of the Milano-Bicocca INFN Section.
The authors acknowledge useful discussions with S. Lepri, R. Livi, F. Ginelli, K. Vynck.
\end{acknowledgments}


\begin{thebibliography}{37}
\bibitem[{ave()}]{average}
\bibinfo{note}{A different average follows from allowing the random walk to
  start at any lattice coordinate, i.e. within the empty regions as well.
  However, it is clear that in the thermodynamic limit the starting sites
  within such regions would dominate the average, and the dynamics would be
  ballistic.}
\end{thebibliography}
\end{document}